\documentclass{article}
\usepackage{amssymb}
\usepackage{graphicx}
\begin{document}
\begin{center}

{\Large\bf Bulk viscosity driving the\\[5pt]
acceleration of the Universe\\[5pt]}
\medskip

{ J. C. Fabris\footnote{e-mail: fabris@cce.ufes.br}, S.V.B.
Gon\c{c}alves\footnote{e-mail: sergio@cce.ufes.br} and R. de S\'a
Ribeiro\footnote{e-mail: ribeiro@cce.ufes.br}} \medskip

Departamento de F\'{\i}sica, Universidade Federal do Esp\'{\i}rito Santo,
29060-900, Vit\'oria, Esp\'{\i}rito Santo, Brazil
\medskip

\end{center}

\begin{abstract}
The possibility that the present acceleration of the universe is
driven by a kind of viscous fluid is exploited. At background
level this model is similar to the generalized Chaplygin gas model
(GCGM). But, at perturbative level, the viscous fluid exhibits
interesting properties. In particular the oscillations in the
power spectrum that plagues the GCGM are not present. Possible
fundamental descriptions for this viscous dark energy are
discussed.
\newline
\vspace{0.7cm}
PACS number(s): 04.20.Cv., 04.20.Me
\end{abstract}

\section{Introduction}

The observations of the dynamics of galaxies, cluster of galaxies
and of the supernova type Ia indicate that about $95\%$ of matter
content of the universe is not composed of baryons
\cite{sahni1,pad1}. A fraction of about $1/3$ of this dark
component appears in the agglomerated structures, and it is called
dark matter, while the remaining $2/3$ appears as a smooth
component, driving the acceleration of the Universe, and is called
dark energy. There is a large number of models trying to take into
account the presence of the dark component. The most popular one
is the so-called $\Lambda$CDM \cite{peebles,sahni2}, where dark
energy is represented by the cosmological constant, while dark
matter is composed of {\it WIMPS}, a cold dark matter, composed of
weakly interacting massive particles which must be relics of a
grand unified phase of the universe, like axions. The $\Lambda$CDM
has achieved great success in explaining the observational data
(even if there claims in the opposite sense \cite{lazkoz,berto1}),
but faces at same time many theoretical difficulties, like a huge
discrepancy (of about $120$ orders of magnitude) between the
predicted and the observed values of the cosmological constant
\cite{pad2,carroll}.
\par
Other models are competitive with $\Lambda$CDM, like quintessence
\cite{steinhardt,jerome} and K-essence \cite{mukha}. Another quite
recent proposal is the Chaplygin gas \cite{pasquier,fabris}, which
is phenomenologically represented by a fluid with negative
pressure which varies with the inverse of the density. The
Chaplygin gas model has been generalized by considering that the
pressure, besides to be negative, depends on an arbitrary power of
the inverse of the density. One of the great advantages of the
Generalized Chaplygin gas model (GCGM) is the possibility of
unifying the description of dark energy and dark matter
\cite{viollier,berto2}: a fraction of this exotic fluid can
agglomerate locally, while the other fraction remains a smooth
component. However, criticisms have been addressed to this
proposal mainly due to its behaviour concerning density
perturbations, which exhibits large oscillations in the resulting
power spectrum which do not appear in the observed power spectrum
of mass agglomeration \cite{waga}.
\par
In our point of view, the question of the oscillations in the
matter power spectrum in the GCGM is controversial. The
oscillations in the power spectrum of the GCG are not transferred
to the baryonic power spectrum, and after all, the direct
observation concerns baryons. But, there are claims that the
oscillations in the dark component is reflected in the $\sigma_8$
normalization \cite{waga}. In this sense, it should be interesting
to find a way out to this problem, keeping at same time the
advantages of the GCGM.
\par
In this work we will explore the possibility that the present
acceleration of the universe is due to a kind of viscous fluid. It
is well known that bulk viscosity can generate an acceleration
expansion \cite{murphy}. But, to our knowledge, such possibility
has been investigated only in the context of the primordial
universe, concerning also the search of non singular models
\cite{pimentel,arbab}. We will consider a simple bulk viscosity
model, in the context of the Eckart formalism \cite{eckart}.
Naturally, this approach is phenomenological. Moreover, the Eckart
formalism is not completely consistent, being a truncation of a
causal theory \cite{israel1,israel2}. Hence, everything that will
be developed here must be later complemented by a fundamental
model which can justify our phenomenological approach.
\par
At background level, the description is equivalent to the GCGM:
the viscous dark energy interpolates a matter dominate phase and a
cosmological constant phase. Hence, all observational tests that
concerns the background behaviour (like the supernova type Ia
data) can be immediately used through the analysis already made
for the GCGM \cite{colistete}. However, at perturbative level, new
features appear: the oscillations that plagues the GCG are absent
here for a large range of the parameters. We make a simplified
comparison with the 2dFGRS ignoring for the moment the presence of
baryons. The goal is to show that it is possible qualitatively
reproduce the general features of the mass power spectrum, with
the absence of expressive oscillations.
\par
This paper is organized as follows. In the next section we
describe the viscous model and determine under which conditions it
can account for the dark component of the universe. In section 3,
a perturbative analysis is made, and the predicted power spectrum
is compared with observations. In section 4, we present our
conclusions, with some perspectives to a more fundamental
motivation of this phenomenological model.

\section{Background model}

Let us consider a homogeneous and isotropic universe filled by a
fluid with bulk viscosity. For simplicity, it will be supposed
that the geometry is given by the flat Robertson-Walker metric,
\begin{equation}
ds^2 = dt^2 - a^2(t)[dx^2 + dy^2 + dz^2] \quad ,
\end{equation}
where $a(t)$ is the scale factor. The equations of motion are
\begin{eqnarray}
3\biggr(\frac{\dot a}{a}\biggl)^2 &=& 8\pi G\rho \quad , \\
\dot\rho + 3\frac{\dot a}{a}(\rho + p^*) &=& 0 \quad ,\\
\label{em3} p^* = p - \xi(\rho)u^\mu_{;\mu} &=& p -
3\xi(\rho)\frac{\dot a}{a} \quad .
\end{eqnarray}
We will consider that $p = \beta\rho$ and that the viscosity
coefficient behaves as $\xi(\rho) = \xi_0\rho^\nu$. Hence, the
equations of motion reduce to
\begin{eqnarray}
\label{be1}
3\biggr(\frac{\dot a}{a}\biggl)^2 &=& 8\pi G\rho \quad , \\
\label{be2}
\dot\rho + 3\frac{\dot a}{a}\biggr[(1 + \beta)\rho -
3\xi_0\rho^\nu\frac{\dot a}{a}\biggl] &=& 0 \quad .
\end{eqnarray}
From equation (\ref{be1}), we obtain the relation
\begin{equation}
\frac{\dot a}{a} = \sqrt{\frac{8\pi G}{3}}\rho^{1/2} \quad ,
\end{equation}
so that the equation (\ref{be2}) can be rewritten as
\begin{equation}
\label{ce}
\dot\rho + 3\frac{\dot a}{a}\biggr[(1 + \beta)\rho -
\bar\xi_0\rho^{\nu + 1/2}\biggl] = 0 \quad ,
\end{equation}
where
\begin{equation}
\bar\xi_0 = 3\sqrt{\frac{8\pi G}{3}} \quad .
\end{equation}
The equation (\ref{ce}) admits the solution
\begin{equation}
\rho = \biggr[\frac{\bar\xi_0}{1 + \beta} + \frac{B}{1 +
\beta}a^{r}\biggl]^\frac{1}{\frac{1}{2} - \nu} \quad ,
\end{equation}
where $B$ is an integration constant and $r = 3(\nu - 1/2)(1 +
\beta)$.
\par
In the case of the generalized Chaplygin gas, where there is no
viscosity, the pressure is given by
\begin{equation}
p = - \frac{A}{\rho^\alpha} \quad .
\end{equation}
The relation between the density and the scale factor is given by
\begin{equation}
\rho = \biggr[A + \frac{B}{a^{3(1 + \alpha)}}\biggl]^\frac{1}{1 +
\alpha} \quad .
\end{equation}
The GCGM and the viscosity model coincides, at background level,
if $\beta = 0$ and if
\begin{equation}
\nu = - \biggr(\alpha + \frac{1}{2}\biggl) \quad .
\end{equation}
In this case, initially the Universe behaves as in the matter
dominated phase, becoming later dominated by a cosmological term.
In the case of the viscosity model, the initial phase is
characterized by a domination of a fluid with equation of state $p
= \beta\rho$, when $\nu < 1/2$. For $\nu > 1/2$ the behaviour
sketched above is inverted: initially there is a superluminal
expansion followed by a subluminal expansion. The analysis based
on the supernova type Ia data for the GCGM can be directly
transferred to the dark viscous model, since it depends on the
background only. In \cite{colistete} a extensive analysis of the
GCG parameters has been made. It has been found that the
prediction for the parameter $\alpha$ is $\alpha = -
0.75^{+4.04}_{-0.24}$. Hence, positive values of $\nu$ ("normal"
viscous behaviour) is preferred. But the dispersion is quite
large.

\section{Perturbative study}

In principle, the most interesting situation in the model
described above, in view of the present acceleration of the
universe, concerns the choice $\beta = 0$ and $\nu = -(\alpha +
1/2)$. These choices lead exactly to the same behaviour of the
GCGM for the evolution of the background. However, here we have a
more {\it normal} situation, where the viscosity grows with
density when $0 < \nu < 1/2$. On the other hand, most of the
criticism on the GCGM concerns the fluctuations on the power
spectrum which leads apparently to a $\sigma_8$ normalization that
is not consistent with observation. Hence, in order to verify if
the viscosity model can lead to improvements with respect to the
GCGM, the scalar fluctuations must be studied. This study will be
done here in the synchronous gauge.
\par
In order to perform this perturbative study, the field equations
are rewritten as
\begin{eqnarray}
\label{be1bis} R_{\mu\nu} &=& 8\pi G\biggr(T_{\mu\nu} -
\frac{1}{2}g_{\mu\nu}T\biggl) \quad ,
\\
\label{be2bis}
{T^{\mu\nu}}_{;\mu} &=& 0 \quad , \\
\label{be3bis} T^{\mu\nu} = (\rho + p^*)u^\mu u^\nu -
p^*g^{\mu\nu} \quad &,& p^* = p - \xi(\rho){u^{\mu}}_{;\mu} \quad
.
\end{eqnarray}
The equations (\ref{be1bis},\ref{be2bis},\ref{be3bis}) are
perturbed by introducing
\begin{equation}
\tilde g_{\mu\nu} = g_{\mu\nu} + h_{\mu\nu} \quad , \quad
\tilde\rho = \rho + \delta\rho \quad , \quad \tilde u^\mu = u^\mu
+ \delta u^\mu \quad , \quad \tilde p^* = p^* + \delta p^* \quad ,
\end{equation}
where $g_{\mu\nu}$, $\rho$, $u^\mu$ and $p^*$ are the background
solutions described before, while $h_{\mu\nu}$, $\delta\rho$,
$\delta u^\mu$ and $\delta p^*$ are small perturbations around
them. The synchronous gauge condition $h_{\mu0}= 0$ is imposed. A
long but straightforward calculation leads to the following
perturbed equations:
\begin{eqnarray}
\ddot h + \biggr(2\frac{\dot a}{a} - 4\pi G\xi\biggl)\dot h -
3\biggr(\frac{\dot a}{a}\biggl)^2\biggr[1 + 3\beta -
9\biggr(\frac{\dot a}{a}\biggl)^2\xi'\biggl]\Delta + 8\pi
G\xi\Theta &=& 0 \quad , \\
\dot\Delta + 9\biggr(\frac{\dot
a}{a}\biggl)^2\biggr[\frac{\xi}{\rho} - \xi'\biggl]\Delta +
\biggr( 1 + \beta - 6\frac{\dot
a}{a}\frac{\xi}{\rho}\biggl)\bigg(\Theta - \frac{\dot h}{2}\biggl)
&=& 0 \quad ,\\
\biggr(1 + \beta - 3\frac{\dot
a}{a}\frac{\xi}{\rho}\biggl)\dot\Theta + \biggr[\frac{\dot
a}{a}\biggr(1 + \beta - 3\frac{\dot
a}{a}\frac{\xi}{\rho}\biggl)\biggr(2 - 3\beta + 9\frac{\dot
a}{a}\xi'\biggl) - 3\biggr(\frac{\ddot a}{a} - \frac{\dot
a^2}{a^2}\biggl)\frac{\xi}{\rho}\biggl]\Theta &=& \nonumber\\
- \frac{\nabla^2}{a^2}\biggr[\biggr(\beta - 3\frac{\dot
a}{a}\xi'\biggl)\Delta - \frac{\xi}{\rho}\biggr(\Theta -
\frac{\dot h}{2}\biggl)\biggl] \quad .
\end{eqnarray}
In these equations, the following definitions were used:
\begin{eqnarray}
h = \frac{h_{kk}}{a^2} \quad , \quad \Delta =
\frac{\delta\rho}{\rho} \quad , \quad \Theta = \partial_iu^i \quad
.
\end{eqnarray}
To deduce these equations, the expression for the perturbation of
the effective pressure has been used. From (\ref{be3bis}), the
perturbation in the effective pressure (a crucial aspect for the
results to be present later), is
\begin{equation}
\delta p^* = \biggr(\beta - 3\frac{\dot a}{a}\biggl)\delta\rho -
\xi(\rho)\biggr(\Theta - \frac{\dot h}{2}\biggl) \quad ,
\end{equation}
\par
These equations can be rewritten in terms of the redshift variable
$z = - 1 + 1/a$, where the scale factor has been made equal to
unity today, $a_0 = 1$. Performing also a plane wave expansion in
the perturbed quantities such that
\begin{equation}
\delta(t, \vec x) = \frac{1}{(2\pi)^{3/2}}\int\delta_k(t)e^{i\vec
k.\vec x}\,d^3k \quad , \quad k = \mbox{wavenumber of the
perturbation} \quad ,
\end{equation}
we end up with the following perturbed equations:
\begin{eqnarray}
h'- \frac{1}{2(1 + z)}\biggr[4 - (1 + \beta)Af(z)^{2\nu -
1}\biggl]h &=&\nonumber \\
- 3\frac{f(z)}{1 + z}\biggr[1 + 3\beta - 3(1 + \beta)\nu
Af(z)^{2\nu - 1}\biggl]\Delta + \frac{1 + \beta}{1 +
z}f(z)^{2\nu - 1}\Theta & & \quad , \\
\Delta'- 3\frac{(1 + \beta)(1 - \nu)}{1 + z}Af(z)^{2\nu - 1}\Delta
&-&\nonumber\\ \frac{1 + \beta}{(1 + z)f(z)}\biggr(1 -
2Af(z)^{2\nu - 1}\biggl)\biggr(\Theta - \frac{h}{2}\biggl) &=& 0
\quad ,\\
(1 + \beta)(1 - Af(z)^{2\nu - 1})\Theta' - \biggr\{\frac{1 +
\beta}{1 + z}\biggr[1 - Af(z)^{2\nu - 1}\biggl]\biggr[2 - 3\beta +
3A\nu(1 + \beta)f(z)^{2\nu - 1}\biggl] &+&\nonumber\\
(1 + \beta)Af(z)^{2(\nu -
1)}f'(z)\biggl\}\Theta &+&\nonumber\\
\frac{k^2}{k_0^2}\frac{1 + z}{f(z)}\biggr\{\biggr[\beta - (1 +
\beta)A\nu f(z)^{2\nu - 1}\biggl]\Delta - \frac{1 +
\beta}{3}Af(z)^{2(\nu - 1)}\biggr(\Theta -
\frac{h}{2}\biggl)\biggl\} &=& 0
\end{eqnarray}
The primes mean derivatives with respect to the redshift $z$. The
following definitions were also employed:
\begin{eqnarray}
f(z) &=& \biggr\{A + (1 - A)(1 + z)^{-r}\biggl\}^\frac{1}{1 -
2\nu} \quad , \\
r &=& 3\biggr(\nu - \frac{1}{2}\biggl)(1 + \beta) \quad ,\\
A &=& \frac{3}{1 + \beta}\sqrt{\frac{8\pi G}{3}}\xi_0\rho_0^{\nu -
1/2} \quad ,\\
\end{eqnarray}
The parameter $k_0$ is associated with the Hublle length, $k_0 =
2\pi H_0/c \sim 2\pi/(3h)\times10^{-3}\,Mpc^{-1}$, $H_0$ being the
Hubble's constant. The recent results from the WMAP measurements
of the anisotropy of the cosmic microwave background radiation
indicates $h \sim 0.7$ \cite{wmap}.
\par
We allow the perturbed equations to evolve from $z = 500$ to $z =
0$, where the final spectrum is computed. The initial conditions
are fixed by using the transfer function
\begin{equation}
T(k) = \frac{B\sqrt{k}}{1 + \frac{8}{\Omega}k +
\frac{4.7}{\Omega^2}k^2} \quad .
\end{equation} where $\Omega$ is
the total density fraction, with respect to the critical density,
which in the present case is $\Omega = 1$. The amplitude $B$ can
also be fixed by using the normalization of the anisotropy of CMB.
Following \cite{pad}, we adopt $B = (24\,h^{-1}\,Mpc)^4$. At $z =
0$ we compute the power spectrum
\begin{equation}
P_k = |\delta_k|^2 \quad .
\end{equation}
\par
The spectrum is computed for a large range of values of $k$. The
comparison with the observational results for the power spectrum
of mass agglomeration obtained through the 2dFGRS program is
displayed in figures $1$, $2$ and $3$ for different values of the
parameters $\nu$ and $A$. The main feature is the absence of
oscillations in the power spectrum of the viscous dark fluid. The
absence of oscillations occurs for positive and negative values of
the parameter $\nu$. In principle this seems to be surprising,
since negative values of $\nu$ should correspond to a positive
sound velocity, which should drive oscillations in the power
spectrum. The reason why these oscillations do not appear is that
the spatial gradient term, which drives oscillations or
instabilities depending on its overall sign, is now composed of
three terms, containing not only the density contrast, but the
velocity and the metric perturbations. The presence of this
combination of terms avoids the appearance of strong oscillations
or instabilities. This combination is due to the form of the
effective pressure. Just to compare, the GCGM, at perturbative
level, contains only the density contrast in the spatial gradient
term.
\begin{figure}
\begin{center}
\includegraphics[width=7cm]{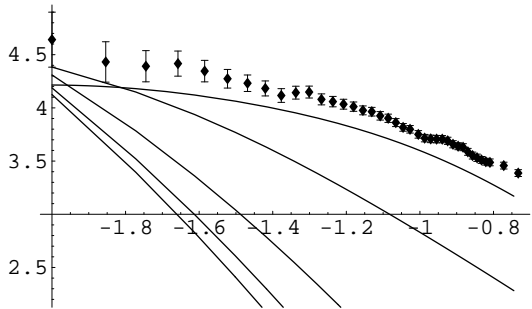}
\end{center}
\caption{Behaviour for $A = 0.1$ and $\nu = - 0.5, 0.3, 0.0, 0.3$
and $0.4$. The ordinate represents $\log_{10}\,P_k$ and the
abscissa $\log_{10}\,kh^{-1}$. As $\nu$ grows, the theoretical
curve approaches the observed curve.}
\end{figure}
\begin{figure}
\begin{center}
\includegraphics[width=7cm]{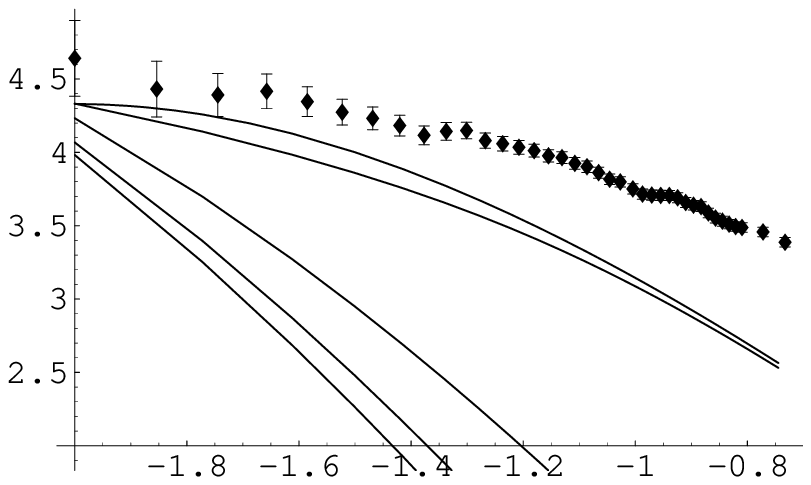}
\end{center}
\caption{Behaviour for $A = 0.4$ and $\nu = - 0.5, 0.3, 0.0, 0.3$
and $0.4$. The ordinate represents $\log_{10}\,P_k$ and the
abscissa $\log_{10}\,kh^{-1}$. As $\nu$ grows, the theoretical
curve approaches the observed curve.}
\end{figure}
\begin{figure}
\begin{center}
\includegraphics[width=7cm]{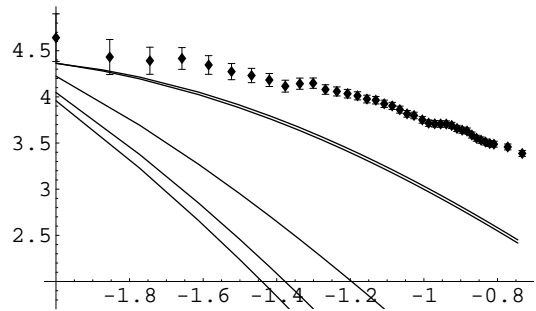}
\end{center}
\caption{Behaviour for $A = 0.7$ and $\nu = - 0.5, 0.3, 0.0, 0.3$
and $0.4$. The ordinate represents $\log_{10}\,P_k$ and the
abscissa $\log_{10}\,kh^{-1}$. As $\nu$ grows, the theoretical
curve approaches the observed curve.}
\end{figure}
\par
The inspection of figures $1$, $2$ and $3$ reveals that the
theoretical curves approach the observational data when $\nu
\rightarrow 1/2$ and $A$ becomes small. A quite reasonable
agreement is, for example, obtained when $\nu = 0.4$ and $A = 0.1$
(figure $1$). Even if the fittings displayed in figures $1$, $2$
and $3$ in general do not reveal a remarkable agreement between
theory and observation, except in the limits stated above, this is
not a serious problem in present context due to one fundamental
reason: we have not considered baryons. In fact, what we should
compute is the power spectrum for the baryonic component, perhaps
with a bias factor which may take into account a contribution of a
fraction of the viscous dark fluid. But, what we would like to
stress is that there is no blow up in the perturbations of this
viscous dark fluid, as it happens with, for example, the GCGM.
Notice that there is a significant suppression of power in the
spectrum for negative values of $\nu$; this suppression is much
less important for positive $\nu$. Such suppression may be
interesting if we remember that we are computing the power
spectrum of the dark component, since the dark component does not
agglomerate completely. Hence, the suppression of power in the
dark component may avoid serious discrepancies with the dynamics
of clusters of galaxies.

\section{Conclusions}

In this work, we have developed a phenomenological model for dark
energy based on a viscous dark fluid. The approach is very
simplified since we consider the bulk viscosity in the Eckart
formalism, ignoring consequently problems of causality. It has
been showed that, with the hypothesis that the bulk viscosity
depends on a power of the density, $\xi = \xi_0\rho^\nu$,
interpolation between a matter dominated phase and a cosmological
constant phase is achieved if $\nu < 1/2$. Hence, such
interpolation can be obtained for a non exotic viscous fluid where
the viscosity decreases with the decreasing of density. Moreover,
the behaviour characteristic of the Chaplygin gas model is
recovered for negative values of $\nu$.
\par
The evolution of density perturbations for this viscous dark fluid
has been computed. There is no oscillation in the power spectrum,
in opposition to what happens with GCGM. The spectrum is highly
suppressed for negative values of $\nu$ but reproduces
qualitatively the observed power spectrum for mass agglomeration
for $\nu$ positive. The fitting of the observational data becomes
quite good when $\nu \rightarrow 1/2$ and $A \rightarrow 0$. The
model studied here contains just one fluid, the viscous dark
energy. Hence, we can expect that the adding of baryons will allow
to fit reasonably the observational data. The absence of
oscillations is due to the fact that the spatial gradient of the
pressure presents a competition between all perturbed quantities.
This is dictated by the covariant representation of the bulk
viscosity.
\par
The phenomenological approach developed here must of course be
supplemented by a fundamental description of this viscous fluid.
To do this a specific fluid model must be considered, with some
interaction between the particles composing this fluid.
Topological defects (cosmic strings, domain walls, textures) can
lead to cosmological fluids with negative pressure in the perfect
fluid approximation. We can think for example on the evolution of
domains wall with friction in an expanding universe
\cite{villenkin}. However, the effective equation of state for
these objects becomes more complex if interactions are taken into
account. Since interactions are inevitable in a gas of topological
defects, it can be expected that deviations from the simple
perfect fluid approximation can appear. We intend to explore this
possibility.
 \vspace{0.5cm}
\newline
{\bf Acknowledgements:} We thank Fl\'avio G. Alvarenga for his
comments on the text and CNPq (Brazil) for partial financial
support.
\newline

\end{document}